\documentclass[aps,prb,article,twocolumn,showpacs,longbibliography,preprintnumbers,amsmath,amssymb,superscriptaddress]{revtex4-2}

\usepackage{graphicx}
\usepackage{color}
\usepackage{mathtools}
\usepackage{amsmath}
\usepackage{bbold}
\usepackage{verbatim}
\usepackage{physics}
\usepackage{bm}
\usepackage{bbm}
\usepackage{cleveref}

\begin{document}

\title{Topology of chalcogen chains}

\author{Adam Kłosiński}
\email{adam.klosinski@fuw.edu.pl}
\affiliation{\mbox{Institute of Theoretical Physics, Faculty of Physics, 
University of Warsaw, Pasteura 5, PL-02093 Warsaw, Poland}}

\author{Wojciech Brzezicki}
\affiliation{\mbox{International Research Centre MagTop, Institute of Physics PAS, Aleja Lotnik\'ow 32/46, PL-02668 Warsaw, Poland}}

\author{Alexander Lau} 
\affiliation{\mbox{International Research Centre MagTop, Institute of Physics PAS, Aleja Lotnik\'ow 32/46, PL-02668 Warsaw, Poland}}
 
\author{\mbox{Cli\`o E. Agrapidis}}
\affiliation{\mbox{Institute of Theoretical Physics, Faculty of Physics, 
University of Warsaw, Pasteura 5, PL-02093 Warsaw, Poland}}

\author{Andrzej M. Ole\'s}
\affiliation{\mbox{Institute of Theoretical Physics, Jagiellonian University, Prof. Stanis\l{}awa \L{}ojasiewicza 11, PL-30348 Krak\'ow, Poland}}
\affiliation{Max Planck Institute for Solid State Research, Heisenbergstrasse 1, D-70569 Stuttgart, Germany}

\author{Jasper van Wezel}
\affiliation{Institute for Theoretical Physics Amsterdam and Delta Institute for Theoretical Physics,\\
University of Amsterdam, Science Park 904, NL-1098 XH Amsterdam, The Netherlands}

\author{Krzysztof Wohlfeld$\,$}
\affiliation{\mbox{Institute of Theoretical Physics, Faculty of Physics, University of Warsaw, Pasteura 5, PL-02093 Warsaw, Poland}}

\date{\today}

\begin{abstract}
We investigate the topological properties of the helical atomic chains occurring in elemental selenium and tellurium. We postulate a realistic model that includes spin-orbit interaction and show this to be topologically non-trivial, with a topological invariant protected by a crystalline symmetry. We describe the end-states, which are orbitally polarized, with an orbital density modulation strongly peaked at the edge. Furthermore, we propose a simplified model that decomposes into three 
orbital chains, allowing us to 
define a topological invariant protected by
a crystalline symmetry. We contrast this result with recent observations made for the orbital 
Su-Schrieffer-Heeger
model containing a $p$-orbital zigzag chain.
\end{abstract}

\maketitle

\section{Introduction}
\label{sec:introduction}

Topology has become one of the pillars of modern condensed matter research over the past decades \cite{moore2010,hasan2010,Qi2011,hasan2011,tokura2019}. Despite the many possible classes of non-trivial topology identified theoretically \cite{atland1997,Chiu13,Chiu14,Sato14,Shio16}, and the many realisations of non-trivial quantum materials identified experimentally \cite{Kon07,Dzi12,Tan12,Mou12,Wan13,Lv15,Xiao18},
suggested direct implementations of the most basic, prototypical models for topological order are rare. Here,
we 
propose
that elemental chalcogens selenium and tellurium 
naturally implement an extension of one of the most fundamental and most well-known topological models in condensed matter, the Su-Schrieffer-Heeger (SSH) model \cite{heeger1988}.

The particular model realised in the spiral chains of chalcogen crystals consists of three independent copies of a period-three SSH model (SSH-3)~\cite{he2020}. 
This model differs from the standard SSH model and the recently introduced orbital SSH model \cite{St-Jean2017, sun2020}, by having a three-site rather than two-site periodicity and lacking exact particle-hole or chiral symmetry \cite{fukutome1984,shimoi1992,Nak02}. It does, however, have a two-fold rotational symmetry that combines with time-reversal to guarantee robust end states in the gap between conduction and valence band. In the chalcogens, the particular charge and orbitally ordered ground state furthermore causes the topological end states to carry a non-zero orbital polarization.

The elemental chalcogens have been studied for almost a century \cite{vonHippel1948,reitz1957,olechna1965} and have been long known to organise into a chiral crystal structure consisting of weakly coupled 
spiral chains \cite{vonHippel1948,reitz1957,Nak02}. Only recently has the origin of these spiral chains been understood in terms of a combined charge and orbital instability of a simple-cubic parent structure \cite{fukutome1984,Silva2018, silva2018_2}. In this instability, a charge density wave develops among quasi-one-dimensional chains of each of the three possible $p$-orbital orientations. Weak Coulomb interaction between the chains leads to a relative alignment of the charge density waves into an emergent three-dimensional structure consisting of spiral chains. Owing to the orbital origin of the charge density distortions, the spiral structure is also orbitally ordered \cite{Silva2018, silva2018_2,klosinski2021}. Although the form of the orbital order depends on the strength of the Coulomb interaction, 
the emerging spiral lattice distortions and their origin in combined charge and orbital order was recently shown to be robust \cite{klosinski2021}.

The topology of the electronic state in the spiral configuration of elemental chalcogens has not been extensively studied, although there are reports from electronic structure calculations that it undergoes a topological phase
transition from an insulator to a Weyl semimetal under strain \cite{agapito2013,ideue2019}. Additionally, there are recent suggestions that surface states may be identified on crystals of elemental selenium and tellurium \cite{pal2013}, but their topological origin has not yet been conclusively established. The particular charge order in the spiral chains, however, is a direct three-site generalisation of the famous two-site SSH model, while the orbital order is strongly reminiscent of the so-called orbital SSH model \cite{St-Jean2017, sun2020}. Here, we show that the elemental chalcogens are indeed topologically ordered, exhibiting a direct realisation of some of the most fundamental known topological models, and giving rise to orbitally polarised end states. The topological invariant characterising the order and end states coincides with the one-dimensional line invariant appropriate for time reversal symmetric systems with a two-fold crystal symmetry \cite{Lau16}.

This paper is organized as follows: In Sec. \ref{sec:model} we introduce the model for a helical chalcogen chain. We begin by describing the chain geometry (Sec. \ref{subsec:chain-geometry}). Then we define the model in two different bases, making use of the helical symmetry (Sec. \ref{subsec:period-three} and \ref{subsec:period-one}), as well as a simplified version of the model, for which the bond angle $\alpha=90^\circ$ and which we call the cubic model (Sec. \ref{subsec:cubic-intro}). In the final subsection of Sec. \ref{sec:model} we discuss the symmetries of the chalcogen chain (Sec. \ref{subsec:symm}). 
In Sec. \ref{sec:results} we show the spectrum of the open chalcogen chain obtained using Exact Diagonalisation (ED) (Sec. \ref{subsec:ed-full}), where we observe in-gap states. We also give a description of the bulk topological invariant (Sec. \ref{subsec:bulk-topology}) and discuss the bulk-boundary correspondence (Sec. \ref{subsec:bulk-boundary}). In the final subsection of this section we discuss the end states of the chalcogen chain (Sec. \ref{subsec:end-states}). 
In Sec. \ref{sec:discussion} we provide a detailed discussion of the topology of the chalcogen chain, going beyond the invariant defined in Sec. \ref{subsec:bulk-topology}. To that end, in Sec. \ref{subsec:cubic-model} we analyze the cubic model -- a simplified version of the chalcogen chain model, with the bond angle $\alpha=90^\circ$. We show that this simpler model factorizes into three SSH-3
chains, one of which supports end-states. These end states are continuously linked to the end-states of the full chalcogen model, allowing us to relate the topology of the chalcogen chain to that of the 
SSH-3
model. In the final subsection of Sec. \ref{sec:discussion} we offer a comparison between the chalcogen chain model and the orbital-SSH model described in \cite{St-Jean2017, sun2020}. 
Finally, in the Appendices we provide a detailed derivation of the end states in the continuous limit (Appendix \ref{sec:end-states-calc}) and a more detailed analysis of the evolution of the energy spectrum of the chalcogen chain with varying spin-orbit coupling strength (Appendix \ref{sec:soc-chirality-breaking}).

\section{Model}
\label{sec:model}
\subsection{Chain geometry and helical symmetry} \label{subsec:chain-geometry}

We previously introduced an electronic model for the helical chains in elemental chalcogens \cite{klosinski2021}, which contains three $p$-orbitals $\{p_x,p_y,p_z\}$ -- but no spin. Here, we expand this model to include spin and spin-orbit coupling. Following the conclusions of Ref.~\cite{klosinski2021}, we assume the effects of the weak Coulomb interaction to be negligible. The non-interacting, electronic model for a single helix then contains three distinct types of bonds, but it can be mapped onto a model with only a single bond type using the helical symmetry.

\begin{figure}[t!]
    \centering
    \includegraphics[width=1\columnwidth]{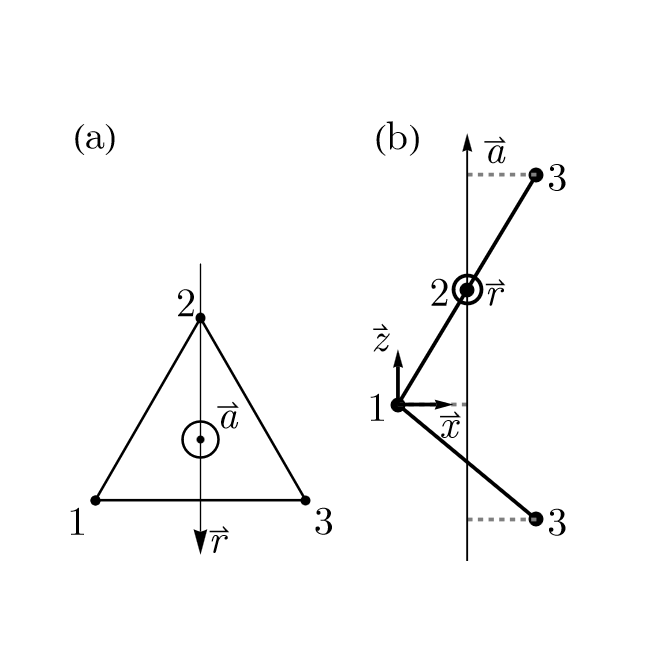}
    \caption{{The geometry of the helical chain: (a) top view, (b)~side view. Marked on the figure: the helical axis ($\vec{a}$), the axis normal to the helical axis ($\vec{r}$), the GO basis vectors $\vec{x}$ and $\vec{z}$ (see Sec. \ref{subsec:period-three} for details) and the shortest distance from each atom to the helical axis (dashed lines).}}
    \label{fig:structure}
\end{figure}

\begin{figure*}[t!]
    \begin{center}
    {\bf \large \hskip -4cm (a) \hskip 8.2cm (b)}\\
    \end{center}
    \includegraphics[width=\columnwidth]{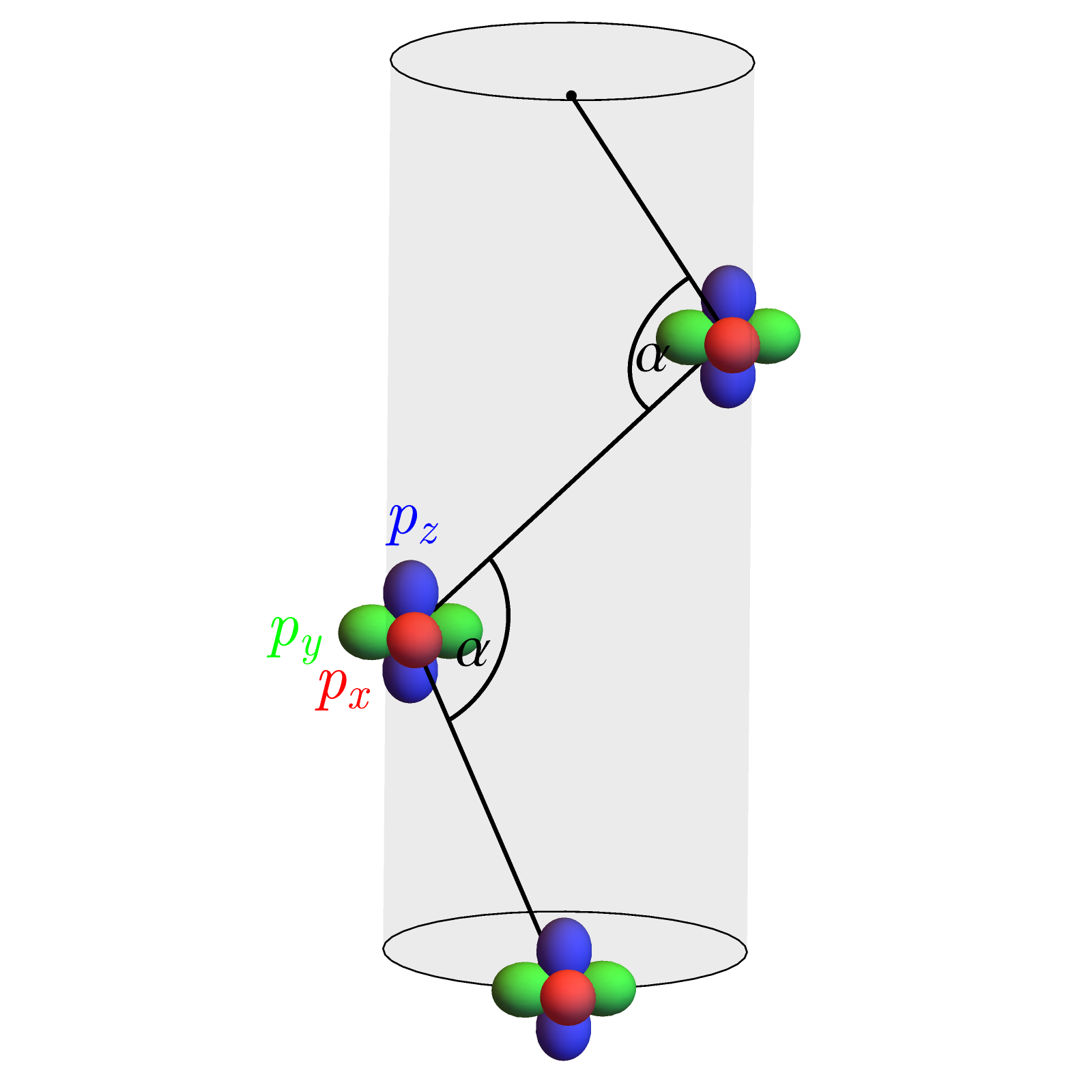}
    \includegraphics[width=\columnwidth]{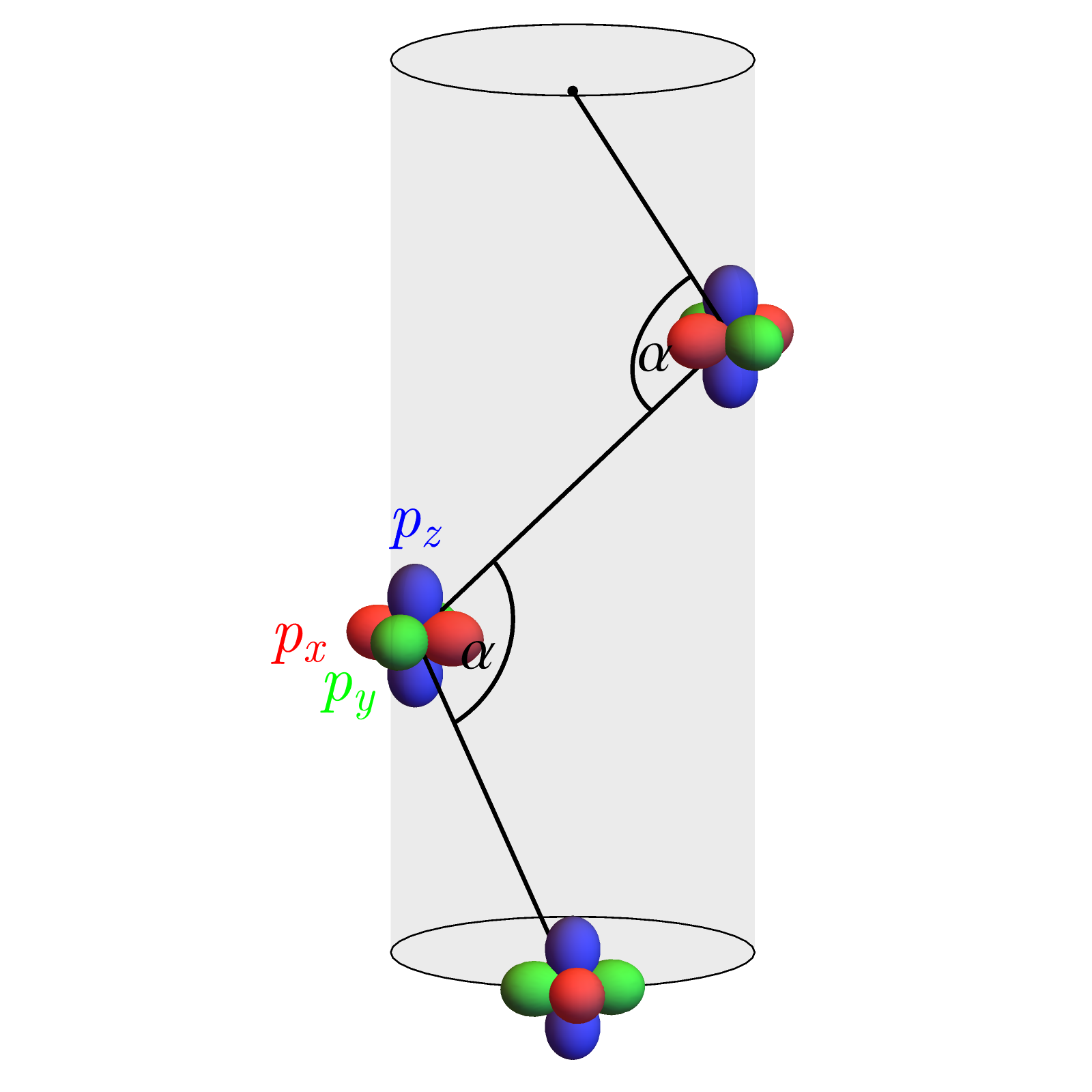}
\caption{The helical chain in three dimensions. The global (a) and the local (b) orbital bases are shown, with $p_x$, $p_y$, and $p_z$ orbitals colored red, green and blue, respectively.
A shaded gray cylinder is also marked, whose axis coincides with the helical axis $\vec{a}$ of the chain, shown in Fig. \ref{fig:structure}. In both cases the bond angle is $\alpha=103^\circ$.}
    \label{fig:chain}
\end{figure*}

Trigonal selenium and tellurium crystals consist of parallel, weakly-coupled helical chains arranged on a hexagonal lattice \cite{reitz1957,olechna1965,matsui2014}. When viewed from the top, atoms in each helix collapse onto the vertices of an equilateral triangle, which demonstrates that the helices are period-three, as shown in Fig.~\ref{fig:structure}. The period-three helix is characterised by a single bond angle $\alpha$ (shown in Fig.~\ref{fig:chain}), which can be treated as a free parameter. For selenium $\alpha_{\text{Se}}\in [102.5^\circ,105.5^\circ ]$, while for tellurium $\alpha_{\text{Te}}\in [102.4^\circ,103^\circ]$. Here we will use $\alpha=103^\circ$, which we call the {\it chalcogen} model, though we will also consider the model for $\alpha=90^\circ$, which we call the {\it cubic} model.

%These models do not have translational invariance $T$ by one bond distance along the helix. However, the helical chain is invariant with respect to a special translation operator $T\otimes O\otimes U$. This symmetry is a combination of: the translation $T$ by one bond distance along the helix {\it times} a 120-degree orbital rotation $O$ {\it times} a 480-degree spin rotation $U$ around the axis of the helix $\vec{a}$ shown in Fig.~\ref{fig:structure}
{These models do not have translational invariance $T$ by one bond distance along the helix. However, the helical chain model is invariant with respect to $T  \otimes O \otimes U $, i.e. a product of the aforementioned translation operator $T$ times a certain orbital rotation $O$ and spin rotation $U$. Here the orbital rotation $O$ is a rotation of orbitals by angle 120-degrees around the axis of the helix $\vec{a}$ -- cf. Fig.~\ref{fig:chain}, which shows the orbital basis before [see panel (a)] and after the orbital rotation [see panel (b)]. Indeed, as Fig.~\ref{fig:structure} (a) suggests, such a rotation, supplemented by the translation by one bond distance along the helix, leaves the sites (and therefore also orbitals) invariant. A bit more subtle situation occurs for the spin rotation $U$ -- here, due to the fact that spins are invariant w.r.t. a 720-degree rotation and not a 360-degree one, the chosen angle of the rotation $U$ has to be $480^\circ$, see [36] for a detailed discussion. Altogether, the system is invariant w.r.t. a combination of: the translation $T$ by one bond distance along the helix {\it times} a 120-degree orbital rotation $O$ {\it times} a 480-degree spin rotation $U$ around the axis of the helix $\vec{a}$ shown in Fig.~\ref{fig:structure}.}
%\footnote{After a 360-degree rotation -- which, for a 120-degree rotation, corresponds to the special translation by a full unit cell -- the spin basis acquires an extra minus sign. Therefore we choose $U$ to be a 480-degree spin rotation. With this choice, we have, on the one hand, $120 = 480 \; \text{mod} \; 360$, and on the other $U^3 = \mathbb{1}$. The first property is essential in order for the spin-orbit interaction to be site-independent\label{footnote}.}.
\footnote{{A spin basis rotation $U$ by the same angle as the orbital basis rotation $O$, that is $120^\circ$, would not work here and instead the rotation angle has to be chosen as $480^\circ$. This is because the fundamental unit cell of the helix has three sites. This means that a translation by three sites $T^3$ is a symmetry. At the same time, if $T \otimes O \otimes U$ is a symmetry, the symmetry applied thrice -- which is $T^3 \otimes O^3 \otimes U^3$ -- is also a symmetry. For both of these operations to be symmetries $O^3$ and $U^3$ need to be identity. While $O^3$ (a $120^\circ$ rotation of the orbital basis applied thrice) is identity, $U^3$ (a $120^\circ$ spin rotation applied thrice) yields an extra minus sign. To satisfy $U^3=\mathbb{1}$ one can try to define $U$ as a $240^\circ$ spin rotation, but then one finds that $O \otimes U$ is not a symmetry of the spin orbit coupling term, and therefore $T \otimes O \otimes U$ is not a symmetry of the Hamiltonian. This is because the angular momentum is rotated by $120^\circ$ and the spin is rotated by $240^\circ$ and consequently their scalar product is not preserved. It is easy to check that the solution is to define $U$ as a $480^\circ$ spin rotation. In this case one has $U^3=\mathbb1$ and the spin-orbit term is invariant with respect to the transformation $O \otimes U$, making $T \otimes O \otimes U$ a symmetry of the Hamiltonian.}\label{footnote}}.
The 120-degree orbital rotation symmetry $O$ is best visible in Figs.~\ref{fig:chain}(a) and \ref{fig:chain}(b), where two orbital bases are shown -- the basis before the orbital rotation [see panel (a)] and the basis after the orbital rotation [see panel (b)].

\subsection{Period-three model} 
\label{subsec:period-three}

%We choose a coordinate system with the $\hat{z}$ axis along the helical axis $\vec{a}$ (see Fig.~\ref{fig:structure}) and the $\hat{x}$ axis in the direction of the shortest line connecting the first site of the helix with its axis.
{We choose a coordinate system with the $\hat{z}$ axis along the helical axis $\vec{a}$ (see Fig.~\ref{fig:structure}) and the
$\hat{x}$ axis in the direction of the shortest line connecting the first
site of the helix with the helical axis $\vec{a}$ (see Fig.~\ref{fig:structure}, this distance is marked for each atom by a dashed line).}
The first site is chosen arbitrarily. We call this basis the global orbital (GO) basis. The GO basis is shown in Fig.~\ref{fig:chain}(a). Since the helix is period-three, the
unit cell consists of atoms at positions $\left(r,0,0\right)$, $\left(-\frac{r}{2},\frac{\sqrt{3}r}{2},a\right)$,
$\left(-\frac{r}{2},-\frac{\sqrt{3}r}{2},2a\right)$, and the translation
vector is $\left(0,0,3a\right)$. On each chain site we have three $p$ orbitals
and spin-orbit coupling, therefore the tight-binding model is given
as follows:
\begin{equation}
H=\sum_{i=1}^{3}T_{i}\otimes h\left(\vec{n}_{i}\right)\otimes\mathbbm{1}_{2}-\lambda\sum_{\alpha=x,y,z}\mathbbm{1}_{N}\otimes L_{\alpha}\otimes\sigma_{\alpha},
\label{eq:ham}
\end{equation}
where $\sigma_{\alpha}$ are the Pauli matrices describing spin and $\left(L_{\alpha}\right)_{\beta\gamma}=-i\varepsilon_{\alpha\beta\gamma}$
are the angular momentum $L=1$ matrices. $\lambda$ is the spin-orbit coupling parameter. 
%Note that the spin-orbit term is negative, because we consider 2/3 filling.
{Note that the spin-orbit term is negative, because we consider the realistic filling for chalcogens -- that is 2/3 \cite{vonHippel1948,reitz1957,klosinski2021} -- and since for more-than-half-filled orbital shells the sign of the spin-orbit coupling term is negative \cite{reis2013}.}
 The $T_{i}$ operators are $N\times N$ matrices which describe nearest-neighbor hopping between sites
of the helix of length $N$ (we assume that $N$ is divisible by three):
\begin{equation}
T_{i}=\sum_{j=1}^{N/3-1}\left(\,\left|i+3j\right\rangle \left\langle i+1+3j\right|+H.c.\,\right),
\end{equation}
with periodic boundary conditions $\left|N+1\right\rangle \equiv\left|1\right\rangle$.
Finally, $h(\vec{n}_{i})$ are the $3\times3$ matrices describing
hopping between $p_{x}$, $p_{y}$ and $p_{z}$ orbitals along the
bonds. They are given by Slater-Koster rules: 
\begin{eqnarray}
h\left(\vec{n}\right)&=&\begin{pmatrix}n_{x}^{2}\delta t+t_{\pi} & -n_{x}n_{y}\delta t & -n_{x}n_{z}\delta t\\
-n_{y}n_{x}\delta t & n_{y}^{2}\delta t+t_{\pi} & -n_{y}n_{z}\delta t\\
-n_{z}n_{x}\delta t & -n_{z}n_{y}\delta t & n_{z}^{2}\delta t+t_{\pi}
\end{pmatrix} \nonumber \\ 
&=& (t_{\pi}-t_{\sigma})(\vec{n}\cdot\vec{L})^2
+t_{\sigma}\mathbbm{1}_{3},
\label{slater}
\end{eqnarray}
where $\delta t=t_{\sigma}-t_{\pi}$ and $t_{\pi}$($t_{\sigma}$)
are the bonding amplitudes of the $\pi$ ($\sigma$) bonds \cite{slater1954}.
Since the helix is period-three, we have three non-equivalent bond directions
denoted as normalized vectors $\vec{n}_{i}$. They can be derived
from the atomic positions as functions of the bond angle $\alpha$. 
They are given by:
\begin{eqnarray}
\vec{n}_{1} & = & \frac{1}{g\left(\alpha\right)}\left(-\frac{3}{2},\frac{\sqrt{3}}{2},\sqrt{3}f\left(\alpha\right)\right),\\
\vec{n}_{2} & = & \frac{1}{g\left(\alpha\right)}\left(0,-\sqrt{3},\sqrt{3}f\left(\alpha\right)\right),\\
\vec{n}_{3} & = & \frac{1}{g\left(\alpha\right)}\left(\frac{3}{2},\frac{\sqrt{3}}{2},\sqrt{3}f\left(\alpha\right)\right),
\end{eqnarray}
with $f\left(\alpha\right)=
\sqrt{\frac{3}{4}\left(\cos\frac{\alpha}{2}\right)^{-2}-1}$,
and $g\left(\alpha\right)=
\frac{3}{2}\left(\cos\frac{\alpha}{2}\right)^{-1}$.
Alternatively, they can be expressed by the ratio of the helix pitch to the distance of the chain atoms from the helical axis $r$, namely $q=a/r$. The value of $q$ is related to the bond angle $\alpha$ by
\begin{equation}
\cos\alpha=\frac{3-2q^{2}}{6+2q^{2}}\,.
\end{equation}
Note that $\frac{\pi}{3}<\alpha<\pi$ while $0<q<\infty$. 

\begin{figure}[t!]
    \centering
    { \hskip 0.8cm \large (a) \hspace{2.1cm} (b)}\\
    \includegraphics[width=1.1\columnwidth]{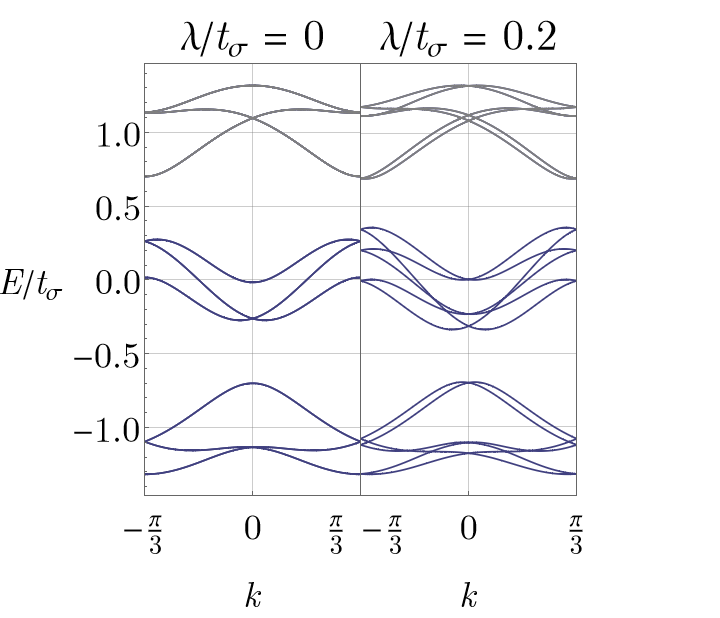}
    \caption{Band structure of the chalcogen model described by Eq.~\eqref{eq:ham}
    (with $\alpha=103^\circ$, $t_\pi / t_\sigma$ = -1/3) with periodic boundary conditions:
    (a) without spin-orbit coupling, $\lambda / t_\sigma = 0$, and (b) with spin-orbit coupling $\lambda / t_\sigma = 0.2$.
    Occupied bands are colored blue, unoccupied bands are colored gray.}
    \label{fig:band-structure}
\end{figure}

The band structure for the model of Eq.~\eqref{eq:ham} is shown in Fig.~\ref{fig:band-structure}. It exhibits band gaps at $1/3$ and $2/3$ filling, which are substantial for realistic values of the spin-orbit coupling parameter $\lambda$, as discussed in Sec.~\ref{subsec:ed-full} below.

\subsection{ Period-one model \hfill} \label{subsec:period-one}
Note that the model of Eq.~\eqref{eq:ham} has a symmetry whenever $N$
is divisible by three described by the operator~$Q$,
\begin{equation} 
\label{eq:q-symm}
Q=T\otimes O\otimes U=T\otimes\exp\left[i\frac{2\pi}{3}L_{z}\right]
\otimes\exp\left[i\frac{8\pi}{3}\frac{\sigma_{z}}{2}\right].
\end{equation}
This represents a cyclic shift by one site along the helix combined with a 120-degree
rotation of orbitals and a 480-degree rotation of spins. $O$ and $U$ are the orbital and spin rotations discussed in Sec.~\ref{subsec:chain-geometry} and $T$ is:
\begin{equation}
T=\left|N\right\rangle \left\langle 1\right| + \sum_{r=1}^{N-1} \,\left|r\right\rangle \left\langle r+1\right|.
\end{equation}
Here, $r$ 
labels the sites along 
the chain.

From $\left[H,Q\right]=0$ and the choice of $N\in3\mathbb{N}$ we
see that in the eigenbasis of $Q$ the Hamiltonian $H$ is decomposed
into $N$ $6\times$6 diagonal blocks. The easiest way to perform
such decomposition, however, is by defining a unitary transformation,
\begin{equation} 
\label{eq:v-trans}
V=\sum_{r=0}^{N-1}\left|r+1\right\rangle \left\langle r+1\right|
\otimes\exp\!\left[i\frac{2\pi r}{3}L_{z}\right]\otimes
\exp\!\left[i\frac{8\pi r}{3}\frac{\sigma_{z}}{2}\right],
\end{equation}
that acts locally in each unit cell. Transforming the Hamiltonian to 
the new basis, which we will call the local orbital (LO) basis, we get
\begin{eqnarray}
H' &\!=\! &\left( T\otimes h\left(\vec{n}_{1}\right)
\exp\!\left[-i\frac{2\pi}{3}L_{z}\right]\!\otimes\!
\exp\!\left[-i\frac{8\pi}{3}\frac{\sigma_{z}}{2}\right]\!+\!H.c.\right)
\nonumber \\
 & - & \lambda\sum_{\alpha=x,y,z}\mathbbm{1}_{N}\otimes L_{\alpha}\otimes\sigma_{\alpha},
\end{eqnarray}
where $H'=VHV^{\dagger}$, with $H$ defined in Eq.~\eqref{eq:ham}.
In the LO basis, the Hamiltonian has a single-site
unit cell. Now we can easily go to $k-$space by inserting the
eigenvalues of $T$ to get:
\begin{eqnarray} \label{eq:ham-k}
H'(k) &\!=\!&\left(\!e^{ik}h\!\left(\vec{n}_{1}\right)
\exp\!\left[-i\frac{2\pi}{3}L_{z}\right]\!\otimes
\exp\!\left[-i\frac{8\pi}{3}\frac{\sigma_{z}}{2}\right]\!+H.c.\!
\right)\nonumber \\
 & - & \lambda\sum_{\alpha=x,y,z}L_{\alpha}\otimes\sigma_{\alpha}.
\end{eqnarray}
Note that this Hamiltonian can also be put in a form that does not 
depend on the choice of the coordinate system by substituting
\begin{eqnarray} \label{eq:coordinate-independent}
L_z & \to & \vec{m}\cdot \vec{L}, \nonumber \\
\sigma_z & \to & \vec{m}\cdot \vec{\sigma}
\end{eqnarray}
in $H'$. In the above, $\vec{m}$ is a normalized vector along the axis 
of the helix and $h\!\left(\vec{n}_{1}\right)$ is already given in
coordinate-independent form in Eq.~\eqref{slater}.

\subsection{The cubic model ($\alpha=90^\circ$)} \label{subsec:cubic-intro}
\begin{figure}[t!]
    \centering
    \includegraphics[width=\columnwidth]{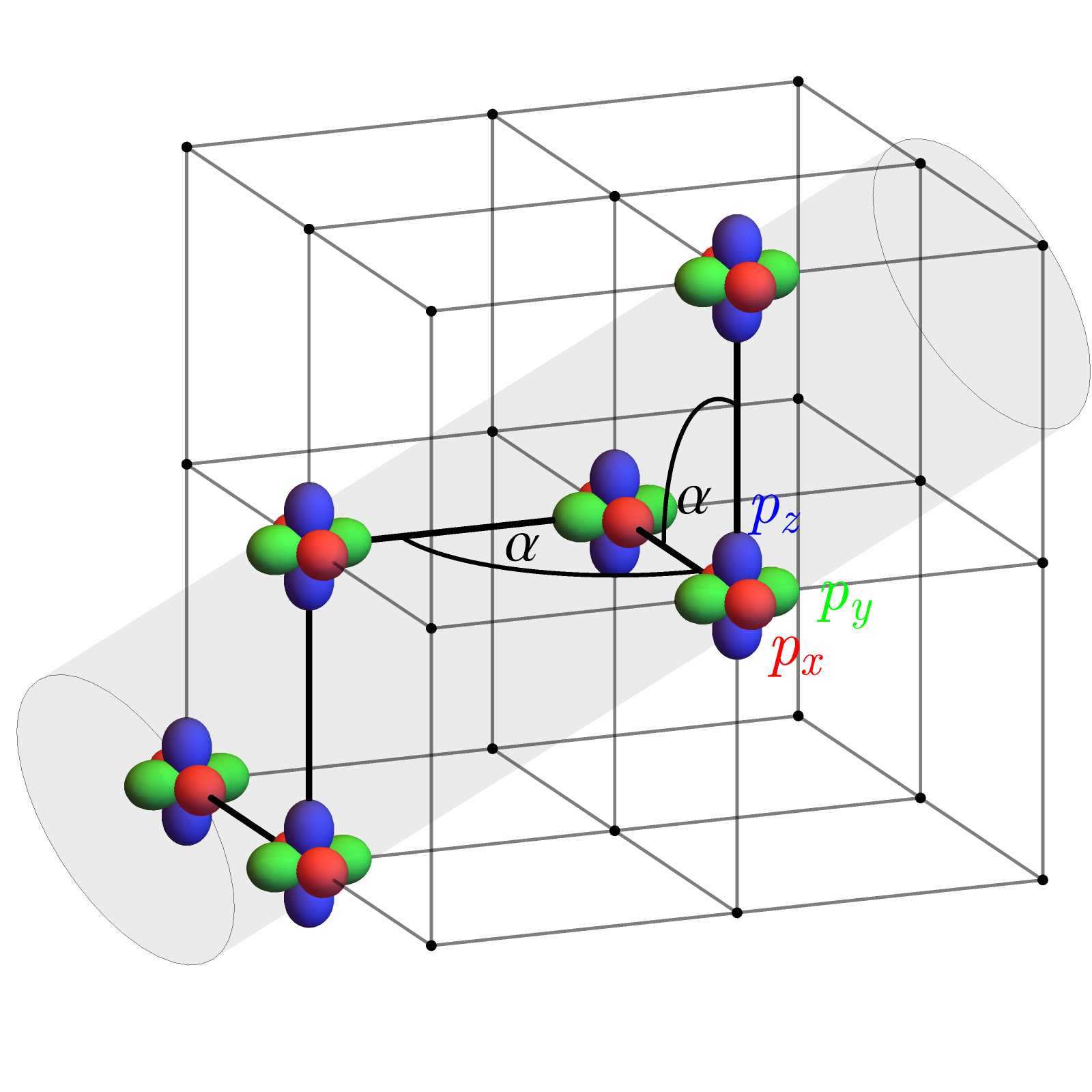}
    \caption{The helical chain for $\alpha=90^\circ$. 
    At each site of the chain the GCO basis is shown (see Sec.~\ref{subsec:cubic-intro}), with $p_x$, $p_y$, and $p_z$ orbitals colored red, green and blue respectively.
    A~shaded gray cylinder is also marked, whose axis coincides with the helical axis $\vec{a}$ of the chain, shown in Fig.~\ref{fig:structure}. The cubic environment is made apparent.}
    \label{fig:cubic-chain}
\end{figure}

The model becomes especially simple for $\alpha = 90^\circ$ 
(the cubic model case), when a different basis choice is convenient.
In the period-1 case we use the coordinate independent form of 
$H'(k)$ and set the direction of the helix as \mbox{$\vec{m}=(1,1,1)/\sqrt{3}$} and of
the first bond as $\vec{n}_1\!=\!(0,0,1)$. This is a valid basis choice for any $\alpha$, and we call this basis choice the local cubic orbital (LCO) basis, as it makes the cubic model Hamiltonian especially simple. In this basis, the orbital hopping part of the Hamiltonian becomes:
\begin{eqnarray}
h\left(\vec{n}_{1}\right)\exp\left[-i\frac{2\pi}{3}\vec{m}\cdot \vec{L}\right]=\begin{pmatrix}0 & t_{\pi} & 0\\
0 & 0 & t_{\pi}\\
t_{\sigma} & 0 & 0
\end{pmatrix}.
\end{eqnarray}

Transforming back to the period-3 model, for $\alpha=90^\circ$ we get Eqs.~\eqref{eq:ham}-\eqref{slater} with
\begin{eqnarray}
\vec{n}_{1} & = & (0,0,1),\\
\vec{n}_{2} & = & (1,0,0),\\
\vec{n}_{3} & = & (0,1,0).
\end{eqnarray}
We call the LCO basis transformed back to the period-three model the global cubic orbital (GCO) basis. It is shown in Fig.~\ref{fig:cubic-chain} for the $\alpha = 90^\circ$ case.

\subsection{Symmetries of the model} \label{subsec:symm}

Besides the $T \otimes O \otimes U$ helical symmetry the Hamiltonian has two other important symmetries:

\smallskip

(i) Time-reversal symmetry $\Theta = i{\cal K}\sigma_y$ ($\cal K$ is complex conjugation), which leads to Kramers degeneracies at time-reversal invariant momenta since $\Theta^2=-1$, and
\begin{equation}
\Theta H'(k) \Theta^{-1} =  H'(-k)^{\star}.
\end{equation}

\smallskip

(ii) $\mathcal{R}$ -- a 180-degree rotation symmetry around the axis $\vec{r}$ normal to the helical axis and going through one of the atoms (see Fig. \ref{fig:structure}). In the LO basis the form of $\mathcal{R}$ is
\begin{equation}
\mathcal{R} = \exp\left[i\pi L_x \right] \otimes \exp\left[i\pi \frac{\sigma_x}{2} \right],
\end{equation}
and it acts on the Hamiltonian as
\begin{equation}
\mathcal{R} H'(k) \mathcal{R}^{-1} =  H'(-k).
\end{equation}
\smallskip
As a further consequence of these two symmetries, there exist an 
anti-unitary operator $\mathcal{V} = \Theta \mathcal{R}$ such that
\begin{equation}
\mathcal{V} H'(k) \mathcal{V}^{-1} = H'(k)^{\star},
\end{equation}
and $\mathcal{V}^2=\mathbb{1}$. Consequently, there exists a basis in which the Hamiltonian is purely real.

\section{Results} 
\label{sec:results}

\subsection{Exact diagonalization of an open chain} 
\label{subsec:ed-full}

\begin{figure}[t!]
    \centering
    \includegraphics[width=\columnwidth]{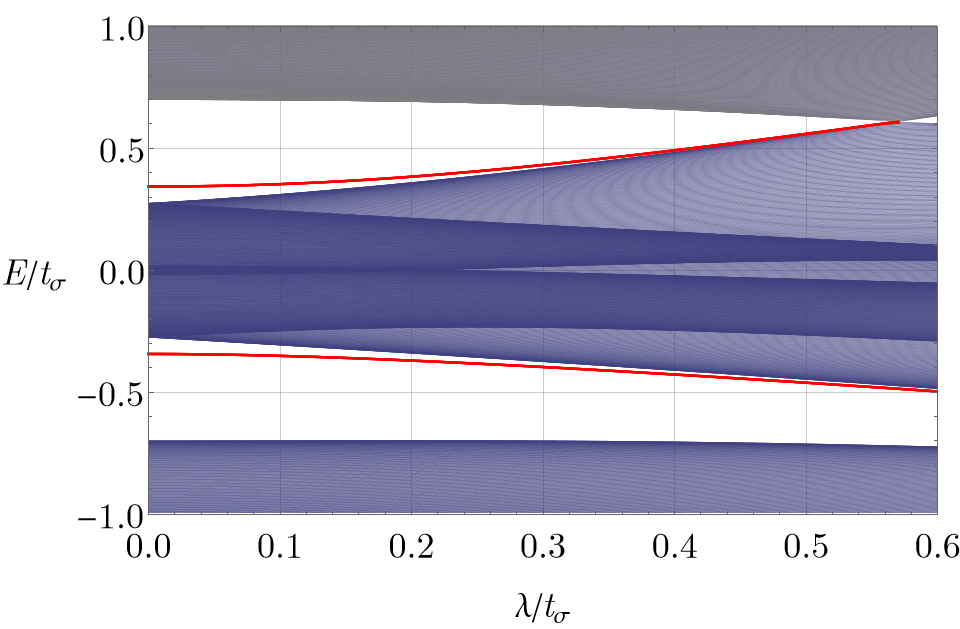}
    \caption{
    The spectrum of
    the chalcogen model in Eq.~\eqref{eq:ham} ($\alpha=103^\circ$, $t_\pi/t_\sigma=-1/3$) on a finite open chain of length $N=501$,
    for different values of $\lambda / t_\sigma$.
    The end states are colored red, the occupied bulk states are 
    colored blue and the unoccupied bulk states are colored gray.}
    \label{fig:ed-full}
\end{figure}

We begin with the exact diagonalisation (ED) of the chalcogen model Hamiltonian in Eq.~\eqref{eq:ham} with open boundary conditions. We use the chain length $N= 501$.
The spectrum is presented in Fig.~\ref{fig:ed-full}. We find that in a finite chalcogen chain with weak spin-orbit interaction four states separated from the continuum appear in each of the two gaps. 
Their presence suggests a nontrivial topology of the chain -- this will be investigated in the following subsections.
%The results of Fig.~\ref{fig:ed-full} also show that as the spin-orbit coupling $\lambda$ increases, the upper gap -- where the Fermi level is located -- closes for $\lambda_{\text{crit.}} / t_\sigma \approx 0.57$ (the value of $\lambda / t_\sigma$ for selenium and tellurium is $0.15 - 0.33$). The in-gap states disappear after the gap closing, indicating a topological phase transition.
{The results of Fig.~\ref{fig:ed-full} also show that as the spin-orbit coupling $\lambda$ increases, the upper gap -- where the Fermi level is located -- closes for $\lambda_{\text{crit.}} / t_\sigma \approx 0.57$. The in-gap states disappear after the gap closing, indicating a topological phase transition. The authors of \cite{herman1963} give the values of the spin-orbit splitting between outermost orbitals, equal to $1.5 \lambda / t_\sigma$, as $0.23$ for selenium and $0.49$ for tellurium. This gives $\lambda / t_\sigma = 0.15$ for selenium and $\lambda / t_\sigma = 0.33$ for tellurium.}

The in-gap states in the lower gap disappear into the bulk at around $\lambda / t_\sigma = 0.8$, while the gap remains open even for very large $\lambda / t_\sigma$. We will disregard the lower gap in our analysis as it lies deep below the Fermi energy. A more detailed discussion of both gaps is presented in Appendix \ref{sec:soc-chirality-breaking}.

\subsection{Bulk topological properties} 
\label{subsec:bulk-topology}

In this subsection we examine the bulk properties and the topological invariant. The system has a spatial symmetry $\cal R$ that can be regarded as a spinful inversion. Therefore, inspired by Ref.~\cite{alexandradinata2014}, we search for the $\mathbb{Z}^{\ge}$ invariant by looking at the spectrum of the Wilson loop operator ${\cal W}$. Its matrix elements can be calculated using a gauge-invariant formula,   
\begin{equation}
{\cal W}_{nm}=\left\langle E_{n}(k_{1})\right|{\cal P}_{n_{f}}(k_{2})\dots{\cal P}_{n_{f}}(k_{N})\left|E_{m}(k_{1})\right\rangle,    
\end{equation}
where $k_j=2\pi j/N$, while $\left|E_{m}(k_j)\right\rangle$ are the eigenstates of $H'(k)$ in Eq.~\eqref{eq:ham-k}, and $N$ should be taken large enough to assure convergence. Notice that only the filled bands are considered in the band labels $m$ and $n$. The operators ${\cal P}_{n_{f}}(k)$ are projectors on the occupied subspace for a given 
$k$:
\begin{equation}
{\cal P}_{n_{f}}(k)=\sum_{n=1}^{n_{f}}\left|E_{n}(k)\right\rangle \left\langle E_{n}(k)\right|.
\end{equation}
Here, $n_f$ is the index of the occupied state corresponding to the largest eigenvalue (energy).

Due to the symmetry $\cal R$, the spectrum of the $\cal W$ operator consists of
the values: $+1$, $-1$ or pairs of complex conjugate numbers of unit modulus \cite{alexandradinata2014}.
The number of $-1$ appearances we denote as $N_{(-1)}$. 
Note that if $\cal R$ was spinless, as in Ref. \cite{alexandradinata2014}, then $N_{(-1)}$ would be equal to the difference in number of occupied states for which ${\cal R} = -1$ between $k=0$ and $k=\pi$ and it would be a topological invariant. In the present case, however, such difference gives zero because in each Kramers doublet we always find two opposite ${\cal R}$ eigenvalues. 

From the general classification involving mirror symmetries \cite{Chiu14}, and more specific work of Ref. \cite{Lau16}, it follows that we can have only one non-trivial phase. Therefore we define a $\mathbb{Z}_2$ topological invariant as
\begin{equation} \label{eq:invariant}
\nu =\frac{1}{2} N_{(-1)} \mod 2.
\end{equation}
This expression is equivalent to the invariant of Ref.~\cite{Lau16}, and is valid as long as $N_{(-1)}$ is even. 
%This is always satisfied because of Kramers degeneracy, which implies that end states in our model come in pairs, and therefore $N_{(-1)}$ has to be even.
{This is always satisfied as the bands come as time-reversal partners and each pair contributes to $N_{(-1)}$ either $0$ or $2$.}
For the chalcogen model we find that the invariant
takes values $\nu=1$ and $\nu=0$ at $2/3$-filling for $\lambda < \lambda_{\text{crit.}}$ and $\lambda > \lambda_{\text{crit.}}$ respectively. 
The critical value $\lambda_{\text{crit.}}$ coincides with the gap closing in Fig.~\ref{fig:ed-full}.

Finally, we would like to point out that in 
one dimension, the bulk-boundary correspondence appears robustly only when there is a symmetry that binds the end-states to zero energy, so either chirality or particle-hole symmetry. In any other case, end-states that appear in the gap can be shifted up or down in energy by a local potential and forced to join the bulk continuum. There are, however, results indicating that a non-trivial Zak phase of the bulk bands leads to charge accumulation at the ends of the system which sometimes manifests itself as the presence of end states \cite{Rhim17,silva2021}.
In our case, the appearance of the end-states in Fig. \ref{fig:ed-full} is fully consistent with the invariant $\nu$ in Eq. \eqref{eq:invariant}. However, in order to better understand what conditions lead to their presence, in the next section we will discuss the bulk-boundary correspondence in our model.

\subsection{Bulk-boundary correspondence} \label{subsec:bulk-boundary}

In order to establish the robustness of the end states, we performed an expansion of the model defined by Eq.~\eqref{eq:ham-k} around the point of the upper gap closing. There are two values of pseudo-momentum $k$ for which the gap closes (see Fig.~\ref{fig:band-structure}), symmetric with respect to the point $k=0$. For now we will focus on the gap closing at pseudo-momentum $k_0>0$. 

In order to expand the Hamiltonian around the gap closing point, we first apply a basis change $u_{k_0,\lambda_\text{crit.}}$ to the eigenbasis of the chalcogen model Hamiltonian in Eq.~\eqref{eq:ham-k} at $k=k_0$ and $\lambda=\lambda_{\text{crit.}}$. 
Next, we apply a projection $P$ of the resulting Hamiltonian onto the two eigenstates forming the Dirac cone at $k_0$ to obtain an effective $2\times2$ Hamiltonian
\begin{equation}
h(k,\lambda) = P u_{k_0,\lambda_\text{crit.}} H'(k)\, u_{k_0,\lambda_\text{crit.}}^\dag P.
\end{equation}
By doing this we find that the eigenbasis of the model at the gap-closing point $(k_0, \lambda_\text{crit.})$ is exactly the basis in which the entire Hamiltonian $u_{k_0,\lambda_\text{crit.}} H'(k) u_{k_0,\lambda_\text{crit.}}^\dag$ is real and whose existence is guaranteed by symmetry (see Sec. \ref{subsec:symm}). Consequently, $h(k,\lambda)$ is also real and we can write it as
\begin{equation} \label{eq:scalar-term}
h(k,\lambda) = s(k-k_0,\lambda) \sigma_0 + f(k-k_0,\lambda) \sigma_x + g(k-k_0,\lambda) \sigma_z,
\end{equation}
where $s(0,\lambda_\text{crit.})=E_0$ (the gap closing energy) and $f(0,\lambda_\text{crit.}) = g(0,\lambda_\text{crit.}) = 0$.
After subtracting the scalar term, the effective $2\times2$ Hamiltonian
\begin{equation}
\tilde{h}(k,\lambda)=h(k,\lambda) - s(k-k_0,\lambda) \sigma_0,
\end{equation}
becomes chiral.

The standard bulk-boundary correspondence guarantees that the chiral Hamiltonian $\tilde{h}(k,\lambda)$
leads to zero-energy end states upon interfacing with a
trivial phase. For $h(k,\lambda)$, however, this is not necessarily the case. This is due to the fact that the scalar term, while irrelevant for the topology of the bulk, does not commute with $\tilde{h}(k,\lambda)$ on an open chain and can therefore affect the end states. Nevertheless, a detailed calculation shows that in this case the end states survive in the gap (see Appendix \ref{sec:end-states-calc} for details).

\subsection{Orbital polarization of the end states} 
\label{subsec:end-states}

\begin{figure}[t!]
    \includegraphics[width=1\columnwidth]{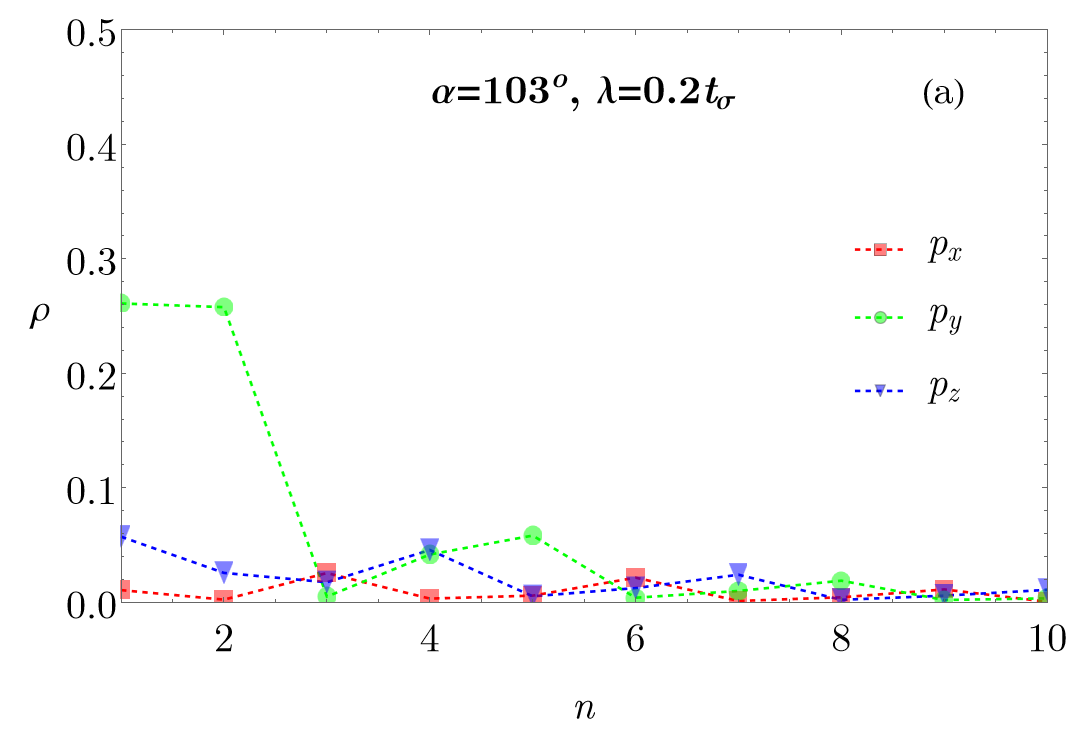}\\
    \includegraphics[width=1\columnwidth]{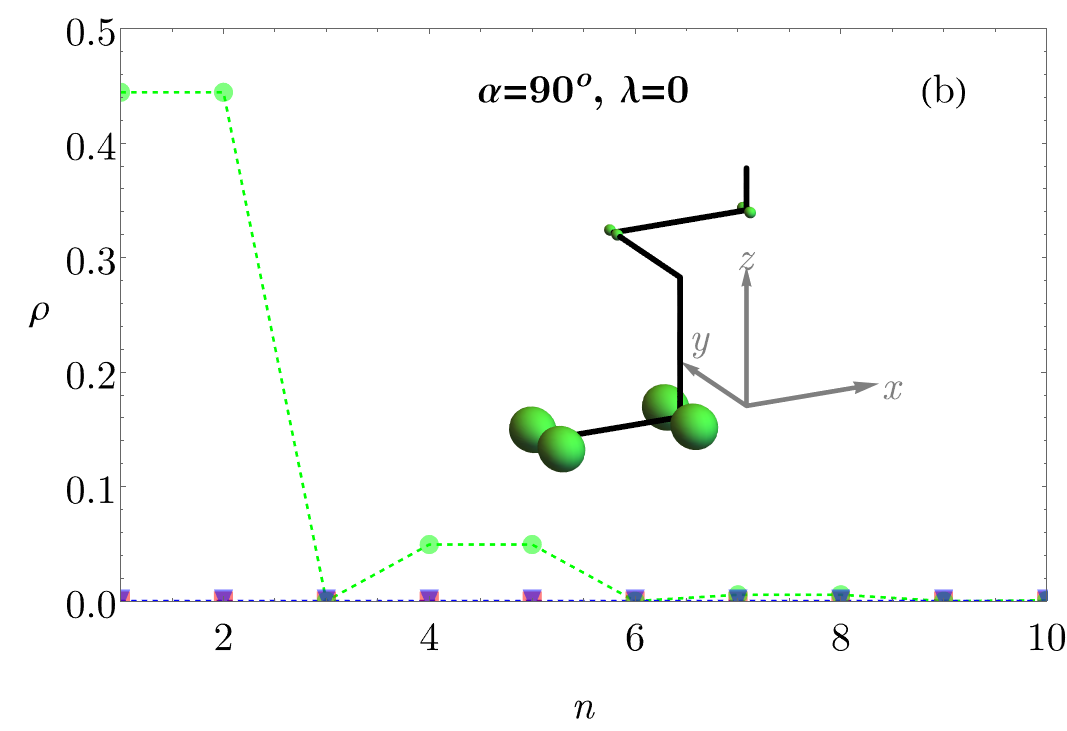}
     \caption{The end-state orbital density for a chain cut across two $y$ bonds: (a) for the chalcogen model of Eq.~\eqref{eq:ham} ($\alpha=103^\circ$, $t_\sigma/t_\pi=-1/3$) with $\lambda / t_\sigma = 0.2 < \lambda_{\text{crit.}} / t_\sigma$, (b) for the cubic model ($\alpha=90^\circ$, $t_\sigma/t_\pi=-1/3$) with $\lambda/t_\sigma = 0$, see Eq.~\eqref{eq:h0} in Sec. \ref{subsec:cubic-model}. In both panels we use the GCO basis defined in Sec. \ref{subsec:cubic-intro} (see Fig. \ref{fig:cubic-chain}). Also shown in panel (b): a schematic picture of the end-state charge density along the helix for the cubic model. The calculation was performed for a chain of length $N=213$ sites.
     }
    \label{fig:end-states}
\end{figure}

We have analyzed the character of the end states by looking at their eigenvectors obtained using ED on an open chain (see Fig.~\ref{fig:end-states}) of size $N=213$. 
The most transparent representation is obtained in the GCO basis.
Looking at Fig.~\ref{fig:end-states}(a), we see that the charge density of the end states decays exponentially away from the edge and that the end states have a strong orbital polarisation. The latter is dictated by the bulk bond pattern and carries a predominantly $p_y$ orbital character.
Since 
in our calculation 
the break in the chain occurs 
where a $p_y$-type dimer would occur in an infinite chain. 

One can identify this result as one closely related to the result 
presented in Fig.~\ref{fig:end-states}(b). It shows the end states of the cubic model ($\alpha =90^\circ$) in the absence of spin-orbit coupling ($\lambda =0$). In this case the end states have a {\it purely} $p_y$ orbital character and the exponentially decaying end state is modulated by a period-three, 1-1-0 charge density wave (CDW). {We have verified that not only in the GCO basis, but indeed in any global orbital basis the perfect orbital polarisation observed in Fig. 6 (b) is absent in Fig. 6 (a).} The physical understanding of this profound result is studied in detail in Sec.~\ref{subsec:cubic-model}.

We conclude that the
spin-orbit interaction and the increase in bond angle 
change 
the nature of the end states only slightly, introducing a small admixture of the other two $p$-orbitals and weakly breaking the 1-1-0 CDW modulation in the $p_y$ orbital density. However, the density of the $p_y$ orbital in the end state is still almost zero on every third site. The non-zero $p_x$ and $p_z$ orbital density is due to the fact that both relevant terms -- the finite spin-orbit coupling $\lambda$ and the deviation from the 90-degree bond angle $\alpha$ -- introduce coupling between the $p$ orbitals, which is absent in the model without these terms. {We note in passing that if one were to further increase $\alpha$, beyond the value $\alpha=103^\circ$ in chalcogens, one would necessarily encounter a topological phase transition. This is because for $\alpha=180^\circ$ the helical chain becomes a simple chain which is not gapped at 2/3 filling and therefore cannot be a topological insulator. However, we have verified that the invariant \eqref{eq:invariant} is robust up to at least $\alpha=120^\circ$, well beyond the bond angle reported for chalcogens.}

\section{Discussion} 
\label{sec:discussion}

\subsection{Cubic model ($\alpha = 90^\circ$) with $\lambda=0$} \label{subsec:cubic-model}   %moved upwards

In what follows we discuss the physical origin of the topological invariant and the orbital polarization of the end states. To this end we turn our attention to the cubic model, which is considerably simpler than the chalcogen model. We uncover a mapping of the cubic model for $\lambda=0$ onto a SSH-3 model with a strong-weak-weak band pattern. In the following subsections we discuss the implications of this finding by showing:\hfill\break
(i) its relevance for the chalcogen model, and\hfill\break
(ii) its relation to the (orbital) SSH model.

We consider a simplified cubic ($\alpha=90^\circ$) model:
\begin{equation} \label{eq:h0}
    H_0 \equiv H \Big|_{\lambda=0, \; \alpha=90^\circ},
\end{equation}
where $H$ is defined in Eq.~\eqref{eq:ham}. Here it is useful to use the 
GCO basis, in which the Hamiltonian takes a particularly simple form (see Sec.~\ref{subsec:cubic-intro} for details). In this basis the $x$, $y$ and $z$ axes point along the bonds of a simple cubic lattice and the helix bonds correspond to consecutive bonds on a simple cubic lattice (see Fig.~\ref{fig:cubic-chain}). 

The Hamiltonian of Eq.~\eqref{eq:h0} separates into two independent spin channels,
\begin{equation} \label{eq:spin-channels}
    H_{0} = \bigoplus_{\sigma} H_0^\sigma.
\end{equation}
Furthermore, each spin block separates into three independent orbital channels, 
\begin{equation} 
    H_0^\sigma = \bigoplus_{\gamma=1}^3 H_0^{\sigma,\gamma},
\end{equation}
each described by a SSH-3 chain. The chains corresponding to different orbitals are shifted by one lattice distance from one another. Their $k$-space Hamiltonians 
(in the GCO basis) are given by,
\begin{equation}
H_0^{\sigma,1}(k) = \begin{pmatrix}0 & t_{\pi} & t_{\sigma}e^{-ik}\\
t_{\pi} & 0 & t_{\pi}\\
t_{\sigma}e^{ik} & t_{\pi} & 0
\end{pmatrix},
\label{H_1}
\end{equation}
\begin{equation}
H_0^{\sigma,2}(k) = \begin{pmatrix}0 & t_{\sigma} & t_{\pi}e^{-ik}\\
t_{\sigma} & 0 & t_{\pi}\\
t_{\pi}e^{ik} & t_{\pi} & 0
\end{pmatrix},
\label{H_2}
\end{equation}
\begin{equation}
H_0^{\sigma,3}(k) = \begin{pmatrix}0 & t_{\pi} & t_{\pi}e^{-ik}\\
t_{\pi} & 0 & t_{\sigma}\\
t_{\pi}e^{ik} & t_{\sigma} & 0
\end{pmatrix}.
\label{H_3}
\end{equation}

\begin{figure}[t!]
    \centering
    \includegraphics[width=0.9\columnwidth]{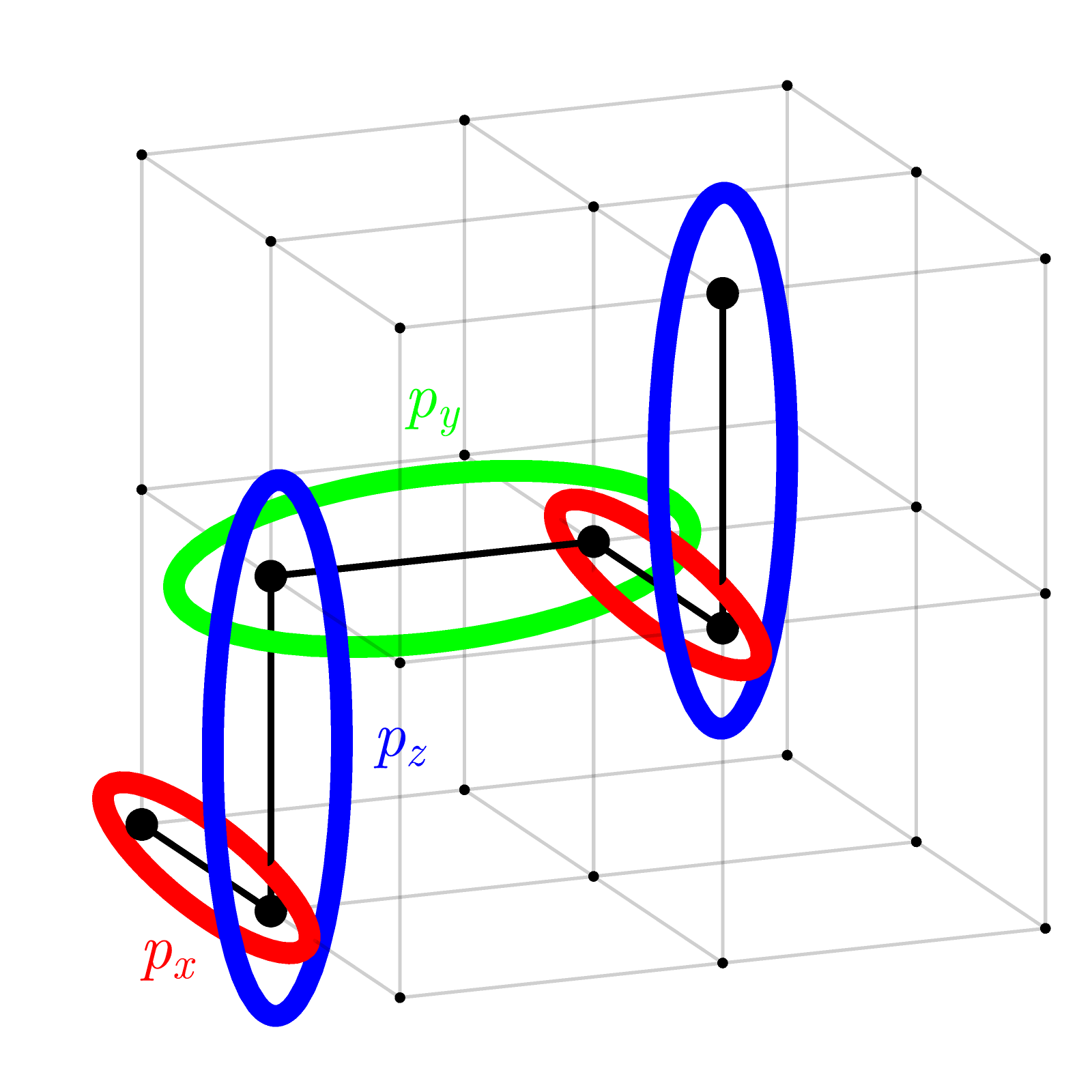}
    \caption{Dimers forming in the cubic model with $\lambda=0$ in Eq.~\eqref{eq:h0}. The helical chain is defined in a cubic environment. Each dimer type, indicated by a different colour, appears on every third bond. For each dimer type a distinct orbital tunneling amplitude is amplified, so that each orbital experiences a $t_\sigma$-$t_\pi$-$t_\pi$ strong-weak-weak tunneling amplitude when moving along the chain. This gives an SSH-3 model for each orbital flavor.}
    \label{fig:3-band-gs}
\end{figure}

Thus, each of the three SSH-3 chains has a strong bond followed by two weak bonds, i.e., a strong-weak-weak bond pattern. This is also visible in the dimer pattern presented in Fig.~\ref{fig:3-band-gs}. Note that the properties of the SSH-3 model are quite different than those of the standard SSH model, as the above Hamiltonians are not chiral-symmetric and are gapless at half-filling. Therefore we cannot get an insulating phase protected by a winding number in contrast to the standard SSH model (see Sec.~\ref{sec:comparison}). 

From the point of view of the crystalline symmetry all three models of Eqs.~\eqref{H_1}, \eqref{H_2}, and \eqref{H_3} are inversion-symmetric, being a manifestation of the $\mathcal{R}$ symmetry present in the helix, but only in the first SSH-3 model the unit cell is compatible with the inversion symmetry taking a form of 
\begin{equation}
{\cal I} = \begin{pmatrix}0 & 0 & 1\\
0 & 1 & 0\\
1 & 0 & 0
\end{pmatrix},
\end{equation}
and yielding
\begin{equation}
{\cal I}H_0^{\sigma,1}(k){\cal I}^{-1} = H_0^{\sigma,1}(-k).
\end{equation}
Now we can use the inversion-invariant of Ref.~\cite{alexandradinata2014} to characterize the insulating phases of $H_0^{\sigma,1}(k)$ at $n=1/3$ and $n=2/3$ filling. The invariant is given by the difference of number of occupied bands with inversion eigenvalue $-1$ between $k=0$ and $k=\pi$. If there are no degeneracy at high-symmetry points, this comes down to:
\begin{equation} \label{eq:invariant-cubic}
\nu' = \sum_{n=1}^{n_f}|\langle E_n(0)\,{\cal I}\,|E_n(0)\rangle 
-\langle E_n(\pi)|\,{\cal I}\,|E_n(\pi)\rangle |,
\end{equation}
where $\nu'$ is a positive-integer topological invariant and 
$n_f$ is the index of the topmost occupied band.
For our choice of $t_\sigma$ and $t_\pi$, we find that $\nu'=1$ both for $n_f=1$ ($n=1/3$ filling) and $n_f=2$ ($n=2/3$ filling). Thus, according to the arguments given in Ref.~\cite{Rhim17}, we can expect that the
open-chain version supports end states, which we discuss below.

Since the models $H_0^{\sigma,\gamma}(k)$ are 
related by a cyclic shift by one lattice site, no matter how we cut the helical chain of the cubic model we should observe a pair of end states (in each spin channel). 
These end states are nothing other than the end states of the inversion-symmetric SSH-3 chain $H_0^{\sigma,1}(k)$. 

By analogy to the ordinary SSH model, it is possible to derive the exact analytic form of the end-states of the model $H_0^{\sigma,1}$. It turns out that their energy is equal to $\epsilon_\pm=\pm t_\pi$, so they satisfy the real-space Schr\"odinger equation of the form
\begin{equation}
(H_0^{\sigma,1} - \epsilon_\pm)|\psi_\pm\rangle = 0,
\end{equation}
with the eigenvectors of the left end states given by
\begin{equation}
\langle i|\psi_-^L\rangle = \begin{pmatrix}\eta & \eta & 0 & \eta^2 
& \eta^2 & 0 & \eta^3 & \eta^3 & 0 &\cdots\end{pmatrix},
\end{equation}
and
\begin{equation}
\langle i|\psi_+^L\rangle = \begin{pmatrix}-\eta & \eta & 0 & \eta^2 
& -\eta^2 & 0 & -\eta^3 & \eta^3 & 0 &\cdots\end{pmatrix},
\end{equation}
where $\eta$
stands for the ratio of the hopping amplitudes, $\eta=|t_\pi/t_\sigma|$.
This shows that these states can be normalized only for $\eta<1$ (this marks the boundary of the topological phase). The right end-states share the same energies and can be obtained from the left ones by applying inversion symmetry $\cal I$, so effectively reversing the order of coefficients of the above vectors.
These states are the exponentially decaying 1-1-0 CDW end states of Fig.~\ref{fig:end-states}(b) 
\footnote{For a finite chain the vectors $\{\ket{\psi^L_{\pm}}\}$ satisfy
\begin{equation} 
\label{eq:edge-state-energies}
\left(H_0^{\sigma,1}-\epsilon_\pm \right) \ket{\psi^L_\pm} = 
\left| \xi_\pm \right>,
\end{equation}
where $\left| \xi\right>$ is a vector with all coefficients but the last equal to 0 and the last coefficient \mbox{$|\bra{N} \ket{\xi}| = |\left(|t_\pi|^{N+1} /|t_\sigma|^N \right)|$} -- exponentially suppressed with increasing system size.}. 
We have verified that $\ket{\psi^L_{\pm}}$ summed over both spin channels agree with the orbital densities shown in Fig.~\ref{fig:end-states}(b), as obtained from exact diagonalization of $H_0$ in Eq.~\eqref{eq:h0}.

\subsection{Relation between cubic and chalcogen models}

\begin{figure}[t!]
    \centering
    \includegraphics[width=\columnwidth]{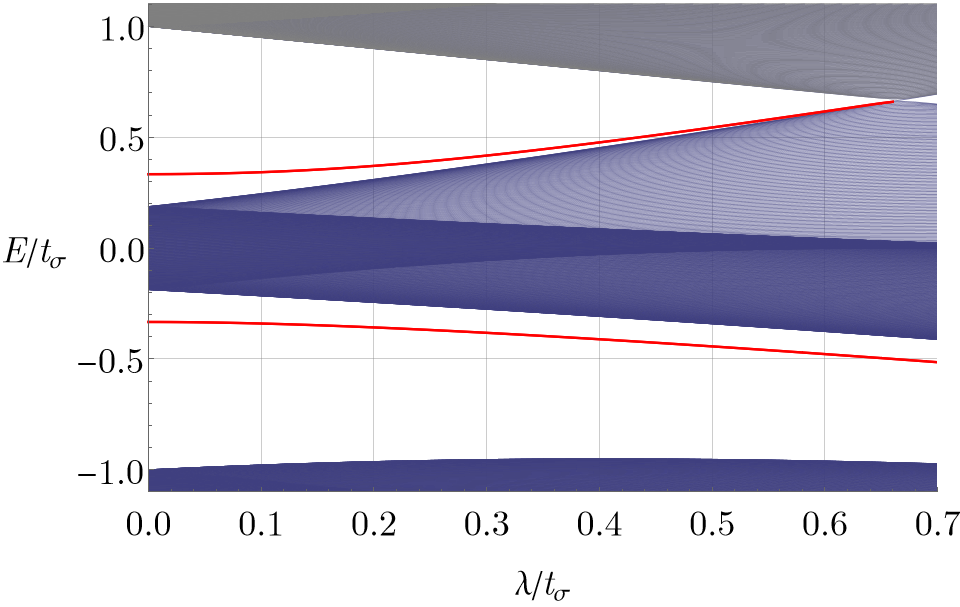}
    \caption{The spectrum of
    the cubic model ($\alpha=90^\circ$, $t_\pi / t_\sigma = - 1/3$) on a finite open chain of Eq.~\eqref{eq:cubic} for chain length $N=501$,
    for different values of $\lambda / t_\sigma$.
    The end states are colored red, the occupied bulk states are colored blue and the unoccupied bulk states are colored gray.}
    \label{fig:3-band-ed}
\end{figure}

We now turn our attention to a more complex model than that in the previous subsection. While we still keep \mbox{$\alpha=90^\circ$}, we consider a {\it finite} value of the spin-orbit coupling $\lambda$. The Hamiltonian of this cubic model is 
\begin{equation} \label{eq:cubic}
    H_{\text{cubic}}(\lambda) \equiv H \Big|_{\alpha=90^\circ},
\end{equation}
where $H$ is given by Eq.~\eqref{eq:ham}.
In Fig.~\ref{fig:3-band-ed} we plot the spectrum of the cubic model \eqref{eq:cubic} on a finite open chain with $\alpha=90^\circ$ as a function of $\lambda/t_\sigma$. The result is very similar to the one obtained for the chalcogen model (Fig.~\ref{fig:ed-full}). For any value of $\lambda$ below the gap closing at $\lambda_\text{crit.} / t_\sigma \approx 0.68$ 
we observe end states.
Moreover, the end states with and without spin-orbit coupling are adiabatically connected in both gaps.

The similarity of Fig.~\ref{fig:3-band-ed} to Fig.~\ref{fig:ed-full}, as well as that of Fig.~\ref{fig:end-states}(a) to Fig.~\ref{fig:end-states}(b), shows that the difference between the realistic chalcogen model of Eq.~\eqref{eq:ham} and the simplified cubic model of Eq.~\eqref{eq:cubic} is purely quantitative. 
%Consequently, we conclude that the bulk-edge correspondence seen in the cubic model of Eq.~\eqref{eq:h0} with $\lambda=0$ (see also Fig.~\ref{fig:3-band-gs}) using the factorization into three SSH-3 models, holds also for small but finite deviations from $\alpha=90^\circ$ bond angle and for $\lambda_{\text{crit.}} > \lambda > 0$. This means that the simple picture of the three SSH-3 models on a helix captures the topological properties of the more realistic chalcogen model. Even though the SSH-3 invariant in Eq.~\eqref{eq:invariant-cubic} present in the simpler cubic model without spin-orbit coupling is no longer well-defined in the chalcogen model, there exists a more general invariant given by Eq.~\eqref{eq:invariant} that characterizes the topological phase of both models. The end states in both models are adiabatically linked and the end states of the chalcogen model are most easily explained using the simple physics of the SSH-3 model.
{Consequently, we conclude as follows. While both the cubic model with $\lambda = 0$ [see Eq.~\eqref{eq:h0} and also Fig.~\ref{fig:3-band-gs}] and the chalcogen model \eqref{eq:ham} possess a non-trivial Lau-Brink-Ortix invariant, given by Eq.~\eqref{eq:invariant}, only the cubic model with $\lambda = 0$ has an additional invariant \footnote{{The additional invariant is a consequence of the absence of spin-orbit coupling.}}, given by Eq.~\eqref{eq:invariant-cubic}. This means that these models are not topologically equivalent. Nevertheless, they are linked by their respective end states: one observes that the end states of the chalcogen model [Fig.~\ref{fig:end-states} (a)] remain similar to these of the cubic model with $\lambda = 0$ [Fig.~\ref{fig:end-states} (b)]. This is commonly seen in topological systems, when upon weak symmetry breaking a change in the invariant (by nature discontinuous) is accompanied by only a quantitative change in the end states \cite{Brzezicki2019}. Therefore, the insight gained from studying the end states of the SSH-3 model can be approximately used to understand the end states of the chalcogen model.}

\subsection{Comparison with the orbital SSH model} 
\label{sec:comparison}

\begin{figure}[t!]
    \centering
    \includegraphics[width=0.9\columnwidth]{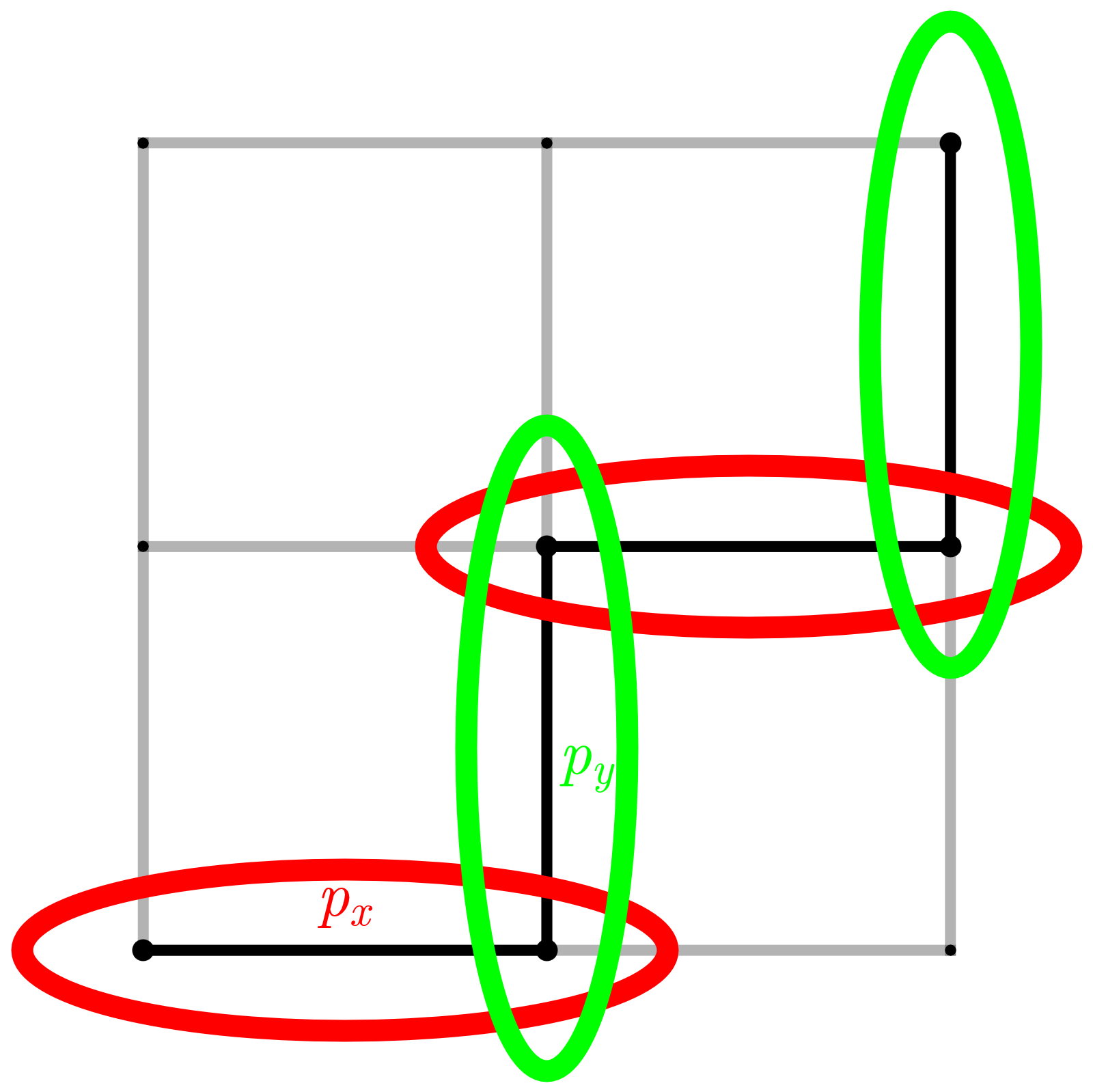}
    \caption{Dimers forming in the orbital SSH model. The zig-zag chain is defined on a simple square lattice. Each dimer type, indicated by a different color, appears on every second bond. For each dimer type a distinct orbital tunneling amplitude is amplified, so that each orbital experiences a strong-weak tunneling amplitude when moving along the chain. This gives an SSH model for each orbital flavor.}
    \label{fig:2-band-model-gs}
\end{figure}

In this section we study electrons in a right-angle zigzag chain on a plane with two active orbitals -- the in-plane $p$ orbitals $p_x$ and $p_y$ \cite{St-Jean2017, sun2020}, which are parallel to the two zigzag bonds. This Hamiltonian has two lattice sites in the unit cell, $a$ and $b$. We consider the filling of 1/2. This model is the orbital SSH model discussed in Refs.~\cite{St-Jean2017, sun2020}, and is given by the following Hamiltonian
\begin{equation} 
\label{eq:orbital-ssh}
    H_{\text{SSH}} = 
    \sum_{i=1}^{2}\tilde{T}_{i}\otimes \tilde{h}_i \otimes\mathbbm{1}_{2},
\end{equation}
where $\tilde{T}_{i}$ operators are $N\times N$ matrices which describe nearest-neighbor hopping between sites
of a zigzag chain of length $N$ (assuming that $N$ is divisible by $2$):
\begin{equation}
\tilde{T}_{i}=\sum_{j=1}^{N/2-1}\left(\,\left|i+2j\right\rangle \left\langle i+1+2j\right|+H.c.\,\right),
\end{equation}
with periodic boundary conditions $\left|N+1\right\rangle \equiv\left|1\right\rangle $. The $2\times2$ matrices $\tilde{h}_i$ describe
hopping between $p_{x}$ and $p_{y}$ orbitals along the two distinct
bonds:
\begin{equation} 
\label{eq:h1}
    \tilde{h}_1 = \left(\begin{array}{cc}
        t_\sigma & 0 \\
        0 & t_\pi 
    \end{array} \right),
\end{equation}
\begin{equation} \label{eq:h2}
    \tilde{h}_2 = \left(\begin{array}{cc}
        t_\pi & 0 \\
        0 & t_\sigma 
    \end{array} \right).
\end{equation}

The orbital SSH model of Eq.~\eqref{eq:orbital-ssh} is topologically non-trivial with four in-gap states. The topological phase is easy to understand by analogy to the SSH model \cite{heeger1988}. First, note that the spin channels separate and one needs to consider only one spin. Second, note that the $\propto t_\sigma$ tunneling processes in Eqs.~\eqref{eq:h1}-\eqref{eq:h2} link $p_x$ orbitals along $x$ bonds and $p_y$ orbitals along $y$ bonds, while the $t_\pi$ tunneling processes link the remaining orbitals on each bond. Since there are no processes mixing orbital flavors, this gives rise to two copies of a half-filled SSH model, one in each orbital channel. The tunneling amplitudes in each channel follow a weak-strong pattern ($\left| t_\pi / t_\sigma \right| < 1$), with the patterns shifted by one lattice distance from each other. The emerging dimer pattern is illustrated in Fig.~\ref{fig:2-band-model-gs}.

In the SSH model we find end states in a finite chain obtained by cutting an infinite chain across two strong bonds (dimers), that is bonds with a larger absolute value of tunneling amplitude -- in our case $t_\sigma$. In other words, the presence of end states is dependent on the cut, i.e., if one cuts across two weak bonds, no end states are found. In Fig.~\ref{fig:2-band-model-gs} we see that no matter how we cut the chain in the orbital SSH model we will always cut two dimers -- it is just the orbital character of these dimers that depends on the cut. We should therefore always find end states in the orbital SSH chain.

The analogy with two decoupled and shifted SSH chains (per spin channel) allows us to understand the four end states in the orbital SSH model. They stem from the bulk invariant -- the winding number of the SSH model. This is in contrast to the cubic model at $\lambda=0$ of Eq.~\eqref{eq:h0} (and, therefore, also the chalcogen model). For the SSH model the invariant comes from the fundamental chiral symmetry and is therefore robust, unaffected by lattice deformations.
In the cubic model, on the other hand, we find an underlying SSH-3 model which is gapless at half-filling and cannot exhibit end-states protected by chiral symmetry.
Instead the end states are protected by the crystalline symmetry (see above). 

\section{Conclusion}
\label{sec:conclusions}

In this work, we have studied the topological properties of the single-element chalcogen crystals selenium and tellurium.
To this end, we have concentrated our efforts on investigating an electronic model with active $p$ orbital degrees of freedom in helical chains -- since such chains form the backbone of the chalcogen crystal structure. While we leave the coupling between the chalcogen chains for a later study, this is not expected to alter the main findings of this work on the existence of topological end states. Rather, the coupling is expected to allow the end states to bundle into coherent surface states. 

Our work has two main results: First,
we have shown that the occupied electronic bands of a realistic model for the chalcogen chains carry a $\mathbb{Z}_2$ topological invariant. The invariant is non-trivial for realistic values of spin-orbit coupling and becomes trivial only for unrealistically large values. 
The invariant is related to the number of particular eigenvalues of the Wilson loop operator $\nu$~\cite{alexandradinata2014}, and is protected by the 180$^{\circ}$ rotation symmetry $\mathcal{R}$ around an axis $\vec{r}$ normal to the helical chain. The end states arising in the topologically non-trivial phase of the open chain are orbitally polarised, i.e., depending on the particular cut of the helical chain, they have a predominantly $p_x$, $p_y$ or $p_z$ orbital character.

Second, we have demonstrated that the onset of the nontrivial topology can be understood in terms of a simplified cubic model. This model also describes a helical chain with active $p$ orbital degrees of freedom, but, unlike for the chalcogens, with the bond angle within the helix being 90$^{\circ}$ and with vanishing spin-orbit coupling.
While the topological properties of the cubic model without spin-orbit coupling are equivalent to the ones found in the realistic chalcogen chain, its topology can also be simply understood as stemming from the topological properties of an SSH-3 model. The SSH-3 model shows a strong-weak-weak bond pattern in one of the three $p$ orbital flavors ($p_x$ or $p_y$ or 
$p_z$), its topological invariant $\nu'$ is protected by the inversion symmetry $\mathcal{I}$, and the end states are entirely orbitally-polarised. 

Note that the SSH-3 model is not only different from the standard
SSH model~\cite{heeger1988}, but also from its orbital variant~\cite{sun2020}. The latter contains a set of two standard SSH chains in each spin channel, one per each $p$-orbital channel.
Furthermore, both the orbital SSH model and the standard SSH model have a winding-number topological invariant protected by chiral symmetry.

\textit{Note added.} -- After this work was completed, we became aware of a recent study of the electronic structure of Se and Te \cite{Zhang}. There, it is suggested that the one-dimensional Se/Te chains are chiral SSH chains. In contrast, we show that the topology of the chalcogen chain stems from the exact Lau-Brink-Ortix invariant~\cite{Lau16}. It can also be understood by analogy to the SSH-3 model, which is topologically distinct from the chiral SSH model. 

\section*{Acknowledgements}

We  kindly  acknowledge  support  by  the  National  Science  
Centre (NCN, Poland) under Projects No. 2016/22/E/ST3/00560, 
2019/34/E/ST3/00404, and 2021/43/B/ST3/02166. The work is 
supported by the Foundation for Polish Science through the 
IRA Programme co-financed by EU within SG OP Programme.
A.L. acknowledges support from a Marie Sk{\l}odowska-Curie 
Individual Fellowship under grant MagTopCSL (ID 101029345).
A.M.O. is grateful for support via the Alexander von Humboldt 
Foundation Fellowship \cite{AvH} (Humboldt-Forschungspreis).

For the purpose of Open Access, the authors have applied a CC-BY 
public copyright licence to any Author Accepted Manuscript (AAM) 
version arising from this submission. 

W.B. and A.L. have contributed equally to this paper.

\begin{appendix}

\section{End states in the continuous limit} 
\label{sec:end-states-calc}

In this section we investigate whether the chirality-breaking scalar term in the chalcogen chain model linearized around the upper gap closing [Eq.~\eqref{eq:scalar-term}] affects the in-gap states.
To this end we use the standard adiabatic argument saying that interfacing a non-trivial
phase of $h(k,\lambda)$ with the vaccum is equivalent to going from the
non-trivial to trivial phase by changing the parameters of the model. This involves passing through the gapless point. The linear expansion
of $h(k,\lambda)$ around the gap closing point is given by
\begin{align}
\begin{split}
h_\text{lin.} &= h(k_{0}+\delta k,\lambda_\text{crit.}+\delta\lambda)  \\
&= E_0 +\delta\lambda \left( a_1 + a_2 \sigma_x + a_3 \sigma_z \right) 
+ \delta k\left(b_1 + b_2 \sigma_z \right)+\\
&+ {\cal O}\left(\delta k^{2},\delta\lambda^{2}\right),
\end{split}
\end{align}
where $a_i$ and $b_i$ are real numbers. To study the real-space interface between trivial and non-trivial phases we work in the continuum setting:
\begin{eqnarray}
\delta k & = & -i\partial_{x},\\ \nonumber 
\delta\lambda & = & \lambda(x),
\end{eqnarray}
and we search for solutions of the Schrodinger equation
of the form:
\begin{equation}
h_{{\rm lin.}}\left(\begin{array}{c}
\psi_{a}(x)\\
\psi_{b}(x)
\end{array}\right)=E\left(\begin{array}{c}
\psi_{a}(x)\\
\psi_{b}(x)
\end{array}\right),
\end{equation}
or equivalently: 
\begin{align}
\begin{split}
& -i \; M \partial_x \left(\begin{array}{c}
\psi_{a}(x)\\
\psi_{b}(x)
\end{array}\right)+R \lambda(x)\left(\begin{array}{c}
\psi_{a}(x)\\
\psi_{b}(x)
\end{array}\right) \\
&=  (E-E_0) \left(\begin{array}{c}
\psi_{a}(x)\\
\psi_{b}(x)
\end{array}\right),\label{eq:eqdiff}
\end{split}
\end{align}
with
\begin{equation}
M=\left(\begin{array}{cc}
b_1+b_2 & 0\\
0 & b_1-b_2
\end{array}\right),
\end{equation}
and
\begin{equation}
R=\left(\begin{array}{cc}
a_1+a_3 & a_2\\
a_2 & a_1-a_3
\end{array}\right).
\end{equation}
We look for solutions of the homogenous equation, that is we look for a state whose energy is equal to the gap closing energy $E_0$. We use the ansatz
\begin{eqnarray}
\psi_{a}(x)=C_a \exp\left(\gamma \int\lambda(x)dx\right) \\
\psi_{b}(x)=C_b \exp\left(\gamma \int\lambda(x)dx\right).
\end{eqnarray}
The homogenous equation now reads:
\begin{equation}
\lambda(x) \exp\left(\gamma \int\lambda(x)dx\right) A \left(\begin{array}{c}
C_{a}\\
C_{b}
\end{array}\right) = 0,
\end{equation}
with
\begin{equation}
A=\left(\begin{array}{cc}
-i b_+ \gamma + a_+  & a_2 \\
a_2 & -i b_- \gamma + a_-) 
\end{array}\right).
\end{equation}
If $\lambda(x)\neq 0$, a solution exists for such $\gamma$ that \mbox{$\text{Det}(A)=0$.} In the case of the chalcogen chain with realistic parameters, that is $\alpha=103^\circ$, $t_\sigma = 1$, and $t_\pi / t_\sigma = -1/3$, we found two solutions: $\gamma_{1} = -0.582 + 0.056 i$, and $\gamma_{2} = 0.582 + 0.056 i$, i.e., $\gamma_{2} = - \gamma_{1}^*$.

Depending on the shape of the domain wall, only one of these solutions is physical and we can set the other to zero so that $\Psi_{i}(\pm\infty)=0$.
For instance, choosing
\begin{eqnarray}
\lambda(x) & = & \Lambda\;\tanh\left(\frac{x}{w}\right),\\
\int\lambda(x)dx & = & \Lambda w \log\cosh\left(\frac{x}{w}\right)>0\,,\nonumber 
\end{eqnarray}
where for $x\to-\infty$ we have a topological phase with $\lambda=\lambda_\text{crit.}-\Lambda$,
and for $x\to+\infty$ a trivial phase with $\lambda=\lambda_\text{crit.}+\Lambda$,
we set $C_{a,2} = C_{b,2} = 0$. Then the solution
is
\begin{equation}
\Psi_{1}(x)=\left( \begin{array}{c}
C_{a,1}\\
C_{b,1}
\end{array} \right) \exp\left(\gamma_1 w \Lambda\log\cosh\frac{x}{w}\right).\label{eq:sol}
\end{equation}
We obtain the localization length of the domain-wall state by taking the limit of large $x$,
\begin{equation}
\Psi_{1}(x\to\pm\infty)\propto e^{\left|x\right| \Lambda \gamma_1},
\end{equation}
hence the localisation length is
\begin{equation}
\xi=\frac{1}{\Lambda \left| \text{Re}(\gamma_1) \right|}\,.
\end{equation}
This characterizes one end state. A similar analysis of the other gap closing point, for $k_0' < 0$, yields another. 

Since the homogenous solution describes a state with energy $E_0$ -- that is in the middle of the gap for all values of $\lambda(x) < \lambda_{\text{crit.}}$ --
and whose localization length is finite, the conclusion from
the above calculation is that in the vicinity of the gap closing the localized end state will survive in the gap even in the presence of the scalar term which breaks the chiral symmetry. 
Let us reiterate that this conclusion applies only in the region around the gap closing, where the linearized model is valid. In particular, the pinning of end states to the gap for all $\lambda(x) < \lambda_{\text{crit.}}$ is a feature which is in contrast with the dispersion of the end states in the upper gap observed in the full model (see Fig.~\ref{fig:ed-full}). The reason for this is that the linearized model is a 2-band $p$-orbital model and consequently the spin-orbit interaction in this model is much simpler than that in the full model, namely, 
\begin{equation}
H_{\text{soc},\text{lin.}} \propto -\lambda L_{z} \otimes \sigma_{z}.   
\end{equation}
As a result, the linearized model does not capture the evolution of the end-state energy. It does, however, show that the end states survive near the gap closing, and taken together with the results of exact diagonalization (see Fig.~\ref{fig:ed-full}) it allows us to conclude that the end states do indeed survive for $\lambda(x) < \lambda_{\text{crit.}}$.

\section{DOS as a function of $\lambda$} \label{sec:soc-chirality-breaking}

\begin{figure}[t!]
\includegraphics[width=\linewidth]{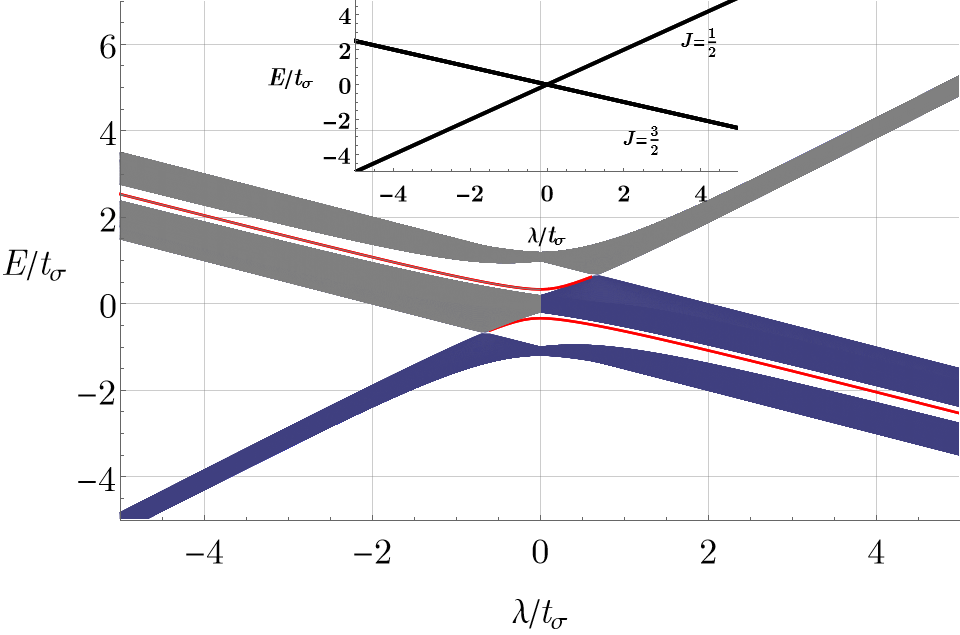}
\caption{
%The spectrum of the cubic model of Eq.~\eqref{eq:cubic} \mbox{($\alpha=90^\circ$, $t_\sigma/t_\pi=-1/3$)} for both positive and negative values of $\lambda / t_\sigma$. The end states are colored red, the occupied bulk states are colored blue and the unoccupied bulk states are colored gray. The $\lambda / t_\sigma < 0$ part of the spectrum corresponds to the filling 1/3, while the $\lambda / t_\sigma \geq 0$ part of the spectrum to the filling 2/3 -- as indicated by the coloring of bands. The inset shows the spin-orbit coupling eigenvalues in the $p$ shell. A symmetry is visible in the DOS upon changing the sign of the spin-orbit interaction, the filling from 1/3 to 2/3 or vice versa, and performing an electron-hole transformation.
{The spectrum of the cubic model of Eq.~\eqref{eq:cubic} \mbox{($\alpha=90^\circ$, $t_\sigma/t_\pi=-1/3$)} for both positive and negative values of $\lambda / t_\sigma$. The end states are colored red, the occupied bulk states are colored blue and the unoccupied bulk states are colored gray. The $\lambda / t_\sigma < 0$ part of the spectrum corresponds to the filling 1/3, while the $\lambda / t_\sigma \geq 0$ part of the spectrum to the filling 2/3 -- as indicated by the coloring of bands [for more on the link between the sign of $\lambda$ and the filling see discussion in Sec. \ref{subsec:period-three}, under Eq. \eqref{eq:ham}]. The inset shows the spin-orbit coupling eigenvalues in the $p$ shell. A symmetry is visible in the DOS upon changing the sign of the spin-orbit interaction, the filling from 1/3 to 2/3 or vice versa, and performing an electron-hole transformation.}}
\label{fig:soc-negative}
\end{figure}

In this Appendix, we provide a simple understanding of the evolution of DOS of the cubic model of Eq.~\eqref{eq:cubic} \mbox{($\alpha = 90^\circ$)} with varying spin-orbit interaction strength $\lambda$. Even though for simplicity we only consider the cubic case, the results of this section are also valid for the chalcogen model of Eq.~\eqref{eq:ham} ($\alpha = 103^\circ$).

An observation can be made about Fig.~\ref{fig:3-band-ed} -- the spectrum of the cubic model on an open chain for $\lambda=0$, given by Eq.~\eqref{eq:h0}, is symmetric with respect to $E/t_\sigma = 0$. However, this model is not chiral --  this is a feature of the open chain alone. 
Because of this, for $\lambda = 0$, the spectrum around 1/3 filling looks the same as the spectrum around 2/3 filling. For $\lambda\neq 0$, however, the physics in the two cases becomes different. While in the upper gap we observe a gap closing, in the lower gap we do not. This asymmetry is the consequence of the fact that spin-orbit interaction breaks particle-hole symmetry. The large $\lambda / t_\sigma$ limit, as well as the gap closing, can be better understood by considering the orbital character of the bands.

The key observation is that, generally, and in particular at the gap closing momentum $k=\pi / 3$, the top and bottom bands have a predominantly $l_z=\pm1$ character, while the middle band has a predominantly $l_z=0$ character, where $l_z$ is the projection of the electron angular momentum onto the $z$ axis of the orbital basis \cite{klosinski2021}. Consequently, the $j=1/2$ pseudospin states
are concentrated mostly in the middle band, while the $j=3/2$ pseudospin states dominate the top and bottom bands. 
Because the spin-orbit 
energy is positive for the $j=1/2$ pseudospin states and negative for the $j=3/2$ pseudospin states, we observe an upwards trend in the energy evolution of the middle band, which is not observed in the top and bottom bands. 
What follows is that there occurs a crossing of the two top bands but no
crossing of the bottom two bands, a features which is visible in 
Fig. \ref{fig:ed-full} and in Fig.~\ref{fig:3-band-ed}.

%Finally, note that if the filling were 1/3 instead of 2/3, the spin-orbit interaction would change sign and the DOS in Figs.~\ref{fig:ed-full} and~\ref{fig:3-band-ed} would flip around the $E/t_\sigma =0$ axis in addition to the Fermi level moving to the lower gap---see Fig.~\ref{fig:soc-negative}. Consequently, the topological properties for the 1/3 filling are the same as for the 2/3 filling.
{Finally, note that if the filling were 1/3 instead of 2/3, the spin-orbit interaction would change sign [see discussion in Sec. \ref{subsec:period-three}, under Eq. \eqref{eq:ham}]. Therefore, the DOS in Figs.~\ref{fig:ed-full} and~\ref{fig:3-band-ed} would flip around the $E/t_\sigma =0$ axis in addition to the Fermi level moving to the lower gap -- see Fig.~\ref{fig:soc-negative}. Consequently, the topological properties for the 1/3 filling are the same as for the 2/3 filling.}

\end{appendix}

\end{document}